\documentclass{article}

\usepackage{arxiv}

\usepackage[utf8]{inputenc} 
\usepackage[T1]{fontenc}    
\usepackage{hyperref}       
\usepackage{url}            
\usepackage{booktabs}       
\usepackage{amsfonts}       
\usepackage{nicefrac}       
\usepackage{microtype}      
\usepackage{lipsum}		
\usepackage{graphicx}
\usepackage{doi}

\usepackage{graphicx}%
\usepackage{multirow}%
\usepackage{amsmath,amssymb,amsfonts}%
\usepackage{amsthm}%
\usepackage{amsmath}
\usepackage{mathrsfs}%
\usepackage[title]{appendix}%
\usepackage{xcolor}%
\usepackage{pgfplots}
\usepackage{tikz}
\usetikzlibrary{positioning,shapes,arrows}
\usepackage{textcomp}%
\usepackage{manyfoot}%
\usepackage{listings}
\usepackage{booktabs}%
\usepackage{algorithm}%
\usepackage{algorithmicx}%
\usepackage{algpseudocode}%
\usepackage{listings}%
\usepackage[backend=biber,style=numeric]{biblatex}
\theoremstyle{thmstyleone}%
\theoremstyle{thmstyletwo}%

\theoremstyle{thmstylethree}%

\title{Formalizing ETLT and ELTL Design Patterns and Proposing Enhanced Variants: A Systematic Framework for Modern Data Engineering}

\author{ 
\href{https://orcid.org/0009-0000-4067-0955}{\includegraphics[scale=0.06]{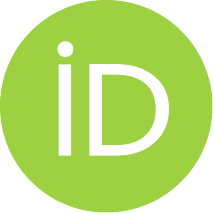}\hspace{1mm}Chiara Rucco}\thanks{Department of Innovation Engineering, University of Salento, Lecce, Italy. Email: \texttt{chiara.rucco@unisalento.it}} \\
Department of Innovation Engineering\\
University of Salento\\
Lecce, Italy \\
\texttt{chiara.rucco@unisalento.it} \\
\And
\href{https://orcid.org/0000-0002-1080-7276}{\includegraphics[scale=0.06]{orcid.pdf}\hspace{1mm}Motaz Saad}\thanks{These authors contributed equally to this work.} \\
Department of Innovation Engineering\\
University of Salento\\
Lecce, Italy \\
\texttt{motazk.saad@unisalento.it} \\
\And
\href{https://orcid.org/0000-0002-6902-0160}{\includegraphics[scale=0.06]{orcid.pdf}\hspace{1mm}Antonella Longo}\thanks{These authors contributed equally to this work.} \\
Department of Innovation Engineering\\
University of Salento\\
Lecce, Italy \\
\texttt{antonella.longo@unisalento.it} \\
}

\renewcommand{\shorttitle}{Formalizing ETLT and ELTL Design Patterns}

\hypersetup{
pdftitle={Formalizing ETLT and ELTL Design Patterns and Proposing Enhanced Variants: A Systematic Framework for Modern Data Engineering},
pdfsubject={q-bio.NC, q-bio.QM},
pdfauthor={David S.~Hippocampus, Elias D.~Striatum},
pdfkeywords={ETL, ELT, Design Patterns, Data Engineering, Cloud, Big Data},
}

\begin{document}
\maketitle

\begin{abstract}
	Traditional ETL and ELT design patterns struggle to meet modern requirements of scalability, governance, and real-time data processing. Hybrid approaches such as ETLT (Extract–Transform–Load–Transform) and ELTL (Extract–Load–Transform–Load) are already used in practice, but the literature lacks best practices and formal recognition of these approaches as design patterns.
This paper formalizes ETLT and ELTL as reusable design patterns by codifying implicit best practices, and introduces enhanced variants, ETLT++ and ELTL++, to address persistent gaps in governance, quality assurance, and observability.
We define ETLT and ELTL patterns systematically within a design pattern framework, outlining their structure, trade-offs, and use cases. Building on this foundation, we extend them into ETLT++ and ELTL++ by embedding explicit contracts, versioning, semantic curation, and continuous monitoring as mandatory design obligations.
The proposed framework offers practitioners a structured roadmap to build auditable, scalable, and cost-efficient pipelines, unifying quality enforcement, lineage, and usability across multi-cloud and real-time contexts. By formalizing ETLT and ELTL, and enhancing them through ETLT++ and ELTL++, this work bridges the gap between ad hoc practice and systematic design, providing a reusable foundation for modern, trustworthy data engineering.
\end{abstract}

\keywords{ETL \and ELT  \and Design Patterns  \and Data Engineering  \and Cloud  \and Big Data}

\section{Introduction}

The exponential growth of data generation and processing demands has dramatically transformed the landscape of data engineering, creating unprecedented challenges for organizations seeking to extract insights from their information assets. With global data generation projected to exceed 394 zettabytes by 2028, traditional data pipeline architectures are increasingly face challenges to meet the complex requirements of modern data-driven enterprises. Data engineering, broadly defined as the discipline of designing and building pipelines to collect, store, and analyze large-scale data, ensures accessibility, reliability, and readiness for analysis \cite{jain2023}. Within this context, design patterns play a crucial role in developing scalable and resilient data architectures capable of handling the massive volumes produced by modern organizations.  

Data Engineering Patterns (DEPs) refer to standardized methods that include ETL processes, data pipelines, and data stream management \cite{DEDPBook2024}. Data Engineering Design Patterns (DEDP) extend this concept by offering best-practice solutions to recurring challenges using optimized and tested approaches \cite{DEDPBook2024}. Among these, the evolution from Extract-Transform-Load (ETL) to Extract-Load-Transform (ELT) has marked a significant shift in data processing, reflecting both technological advances and changing business requirements.  

Traditional ETL processes, which have dominated data warehousing for over two decades, face critical limitations in scalability, real-time processing, and quality assurance. While ETL ensures quality through pre-load transformations, its batch-oriented nature introduces inherent latency that modern applications cannot tolerate. The rise of cloud-native data warehouses catalyzed the transition toward ELT, enabling organizations to exploit massive computational power for in-database transformations \cite{IJFMR2024}. Yet, recent empirical studies document that both ETL and ELT approaches suffer from substantial quality assurance gaps. Foidl et al. \cite{foidl2024pipeline} identified 41 distinct factors influencing pipeline quality, revealing that 78\% of data quality issues originate from insufficient validation at ingestion points. Complementing this, Munappy et al. \cite{munappy2020challenges} show that 67\% of practitioners struggle with integrating heterogeneous sources, while 89\% report persistent consistency issues across distributed environments. These findings converge on a critical observation: data-related issues are primarily caused by incorrect data types and occur predominantly during data cleaning stages.  

Beyond quality, the modern data stack faces five systemic challenges:  
\begin{enumerate}
    \item Fragmented tooling that increases integration overhead 
    \item Operational complexity in orchestration and management
    \item Persistent data quality gaps that consume up to 80\% of engineers’ time
    \item Metadata debt and lineage difficulties that hinder governance.
\end{enumerate}  

While hybrid approaches like Extract-Transform-Load-Transform (ETLT) have been mentioned in technical reports, there is, to the best of our knowledge, no formal academic definition or systematic specification of ETLT as a design pattern. Similarly, the Extract-Load-Transform-Load (ELTL) pattern has not been previously conceptualized in the literature. Existing work predominantly focuses on ETL and ELT in isolation, without considering the systematic formalization of hybrid approaches necessary for complex enterprise environments.  

In practice, ETLT and ELTL patterns have emerged organically, applied by data engineers based on situational needs and personal expertise. Yet, without formal frameworks, standardized definitions, or established best practices, these hybrid approaches lack consistency, reduce reusability, and limit their potential benefits. This paper addresses these gaps by formally defining and analyzing the two hybrid integration patterns, ETLT (Extract, Transform, Load, Transform) and ELTL (Extract, Load, Transform, Load), and by introducing their enhanced variants, ETLT++ and ELTL++. Complementing their formal structure, we provide practical guidance and best practices spanning cloud-agnostic architecture, data quality and validation, performance optimization, error handling and recovery, and monitoring and observability, reflecting concerns repeatedly highlighted in recent literature reviews on extending data integration models for big data and business intelligence.  

The contributions of this paper are threefold:  
\begin{itemize}
    \item Formalization of ETLT and ELTL as distinct hybrid design patterns that reconcile the operational trade-offs of ETL and ELT in contemporary environments, including compute locality, raw data retention, and governance alignment.  
    \item Definition of enhanced ETLT and ELTL design patterns (ETLT++ and ELTL++), a comprehensive best-practices framework covering architecture, data quality, performance, reliability, and observability, grounded in the recurring limitations and requirements surfaced by recent systematic reviews.  
     \item  Implementation artifacts, including structural diagrams and code templates, to accelerate adoption and facilitate rigorous operationalization in multi-cloud and mixed workload contexts, consistent with calls for pattern-driven, reusable data pipeline design in the data engineering literature.

\end{itemize}  

By unifying auditability and agility (via ELTL) with fault isolation and reusability (via ETLT), this work offers data engineering teams a practical roadmap for adopting hybrid patterns that deliver measurable improvements in efficiency, reliability, and cost optimization under modern constraints. The remainder of this paper is organized as follows: 
Section~\ref{stateofart} reviews related work on ETL/ELT, highlighting gaps in the literature and emphasizing the disconnect between academic research and industry practice. Section~\ref{sec:EtlElt} introduces the ETLT and ELTL design pattern structures, which are further extended in Sections~\ref{sec:etltpp} and~\ref{sec:eltlpp}. Finally, Section~\ref{futureworks} discusses future directions, followed a proposal of a benchmarking plan and conclusion.

\section{State of the Art}\label{stateofart}

\subsection{Data Engineering Design Patterns}
Data Engineering Design Patterns (DEDPs) represent a systematic adaptation of classical software design principles to address the specific challenges inherent in large-scale data management and processing systems \cite{gamma1994design, konieczny2024data}. These patterns offer standardized solutions to recurring challenges in building and maintaining robust data infrastructures, encompassing architectural choices, data modeling techniques, and storage/retrieval strategies that aim to create scalable, resilient, and efficient data ecosystems \cite{konieczny2024data, kleppmann2017designing}. Drawing inspiration from the foundational work of the Gang of Four design patterns \cite{gamma1994design}, DEDPs extend beyond traditional object-oriented programming paradigms to address the unique requirements of distributed data systems, real-time processing, and enterprise-scale data governance.

The theoretical foundation of DEDPs builds upon three fundamental categories established in classical software design: creational, structural, and behavioral patterns \cite{gamma1994design}. Creational patterns in data engineering focus on data instantiation and initialization mechanisms, governing how data objects, schemas, and pipeline configurations are generated and managed across distributed systems. Structural patterns define the composition and organization of data components, determining how different data processing modules, storage systems, and integration layers interact to form coherent data architectures \cite{fowler2002patterns}. Behavioral patterns govern the dynamic aspects of data processing workflows, including data flow orchestration, error handling mechanisms, and adaptive responses to changing system conditions \cite{hohpe2003enterprise}.

The evolution of DEDPs reflects the progressive sophistication of enterprise data requirements and technological capabilities. Early patterns focused primarily on Extract-Transform-Load (ETL) processes and batch processing paradigms, suitable for traditional data warehousing environments with predictable data volumes and processing schedules \cite{inmon2005building, kimball2013data}. The emergence of big data technologies and cloud computing platforms necessitated the development of more sophisticated patterns, including the Lambda Architecture for handling both batch and real-time processing \cite{marz2015big}, the Kappa Architecture for stream-centric data processing \cite{kreps2014questioning}, and more recently, the Data Lakehouse pattern that combines the flexibility of data lakes with the transactional consistency of data warehouses \cite{armbrust2021lakehouse}.

Contemporary DEDPs address increasingly complex scenarios involving microservices architectures, event-driven systems, and artificial intelligence integration \cite{microsoft2024patterns}. The microservices-based data pipeline pattern enables organizations to decompose monolithic data processing systems into independent, loosely coupled services that can be developed, deployed, and scaled independently. Event-driven architecture patterns facilitate real-time responsiveness through asynchronous event propagation, enabling immediate data updates and business rule execution \cite{hohpe2003enterprise}. The integration of Large Language Models (LLMs) into data engineering workflows represents an emerging frontier, with patterns beginning to emerge for automated pipeline generation, intelligent data transformation, and semantic data processing.

However, significant challenges remain in effectively applying and adapting DEDPs to specific organizational contexts. Existing research often focuses on individual patterns in isolation, with insufficient attention to the interplay and composition of multiple patterns within complex data architectures \cite{konieczny2024data}. The dynamic nature of data requirements, coupled with the rapid emergence of new technologies, necessitates continuous evolution and adaptation of established patterns. Furthermore, the lack of comprehensive frameworks for pattern selection and composition based on specific business needs and technical constraints represents a critical gap in current knowledge. This thesis addresses these limitations by developing a systematic approach to pattern composition, automated pattern selection methodologies, and the integration of AI-driven capabilities into traditional data engineering design patterns.

\subsection{Foundations of Data Integration: ETL and ELT Paradigms}

The evolution of data integration has been fundamentally shaped by Extract-Transform-Load (ETL) and Extract-Load-Transform (ELT) paradigms, which represent the bedrock of traditional and modern data warehousing architectures respectively. ETL, established as the de facto standard for over two decades, operates on the principle of executing comprehensive data transformations—including cleansing, normalization, validation, and enrichment—prior to loading processed data into target analytical repositories \cite{kimball2002data, abadi2016design, hellerstein2017data}. Traditionally, ETL centralizes and organizes data by extracting it from source systems, applying business logic during transformation, and loading it into a data warehouse and downstream data marts to align with specific departmental needs \cite{reis2022fundamentals}. ELT reverses this sequence by loading raw data directly into the warehouse staging area and leveraging its compute power for in-place transformations, reducing upfront latency and enabling more flexible, cloud-native processing \cite{reis2022fundamentals}. 

\begin{figure}[!ht] \centering \includegraphics[scale=0.23]{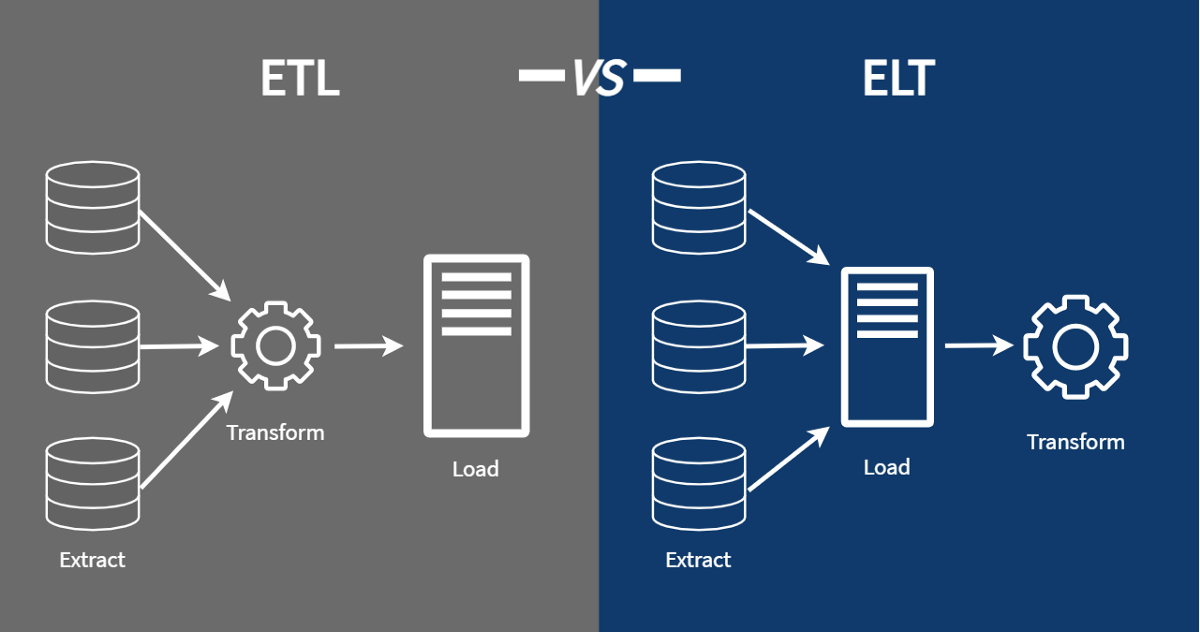} \caption{ETL vs ELT} \label{fig:etlelt} \end{figure}

However, ETL workflows are increasingly challenged by the exponential growth in data volumes and velocity demands of modern enterprises. The batch-oriented nature of ETL processing introduces significant latency—often measured in hours or days—which conflicts with contemporary requirements for near-real-time analytics and responsive business intelligence \cite{burton2017etl, kim2013large}. Moreover, ETL architectures struggle with scalability limitations due to their reliance on dedicated transformation servers and complex orchestration mechanisms, making them unsuitable for the elastic, distributed processing demands of cloud-native environments \cite{IJFMR2024}.

Responding to these limitations, ELT has gained prominence by fundamentally reversing the transformation-loading order and capitalizing on the elastic compute capabilities of modern cloud-native data warehouses \cite{goncalves2019data, rivera2020cloud, IJFMR2024}. ELT facilitates rapid ingestion of raw data into scalable storage systems and defers complex transformations to postload stages, where they are performed in situ using the target system's computational resources. This approach dramatically reduces initial processing delays, enables iterative refinement of transformation logic, and supports more flexible schema evolution \cite{alex2019cloud}.

Despite these advantages, ELT introduces significant challenges that have been extensively documented in both academic literature and by practitioners. The absence of a preliminary validation of the quality of data in ELT can allow flawed, inconsistent, or malformed data to propagate downstream, severely complicating governance efforts and potentially increasing storage costs \cite{schelter2018deequa, burton2017etl}. Furthermore, ELT workflows may suffer from unpredictable query performance, particularly when handling semi-structured or deeply nested datasets common in IoT, event stream processing, and modern application architectures \cite{kimball2020developing, munappy2020challenges}. 

It can be noted from the previous review that while ETL and ELT have been extensively studied, the academic treatment of hybrid approaches, such as ETLT and ELTL patterns, is particularly limited. Scattered references to these paradigms appear in industry whitepapers and technical blogs \cite{macrina2018structured, pace2019lineage}, but lack the rigorous formal definition, comparative analysis, and systematic evaluation necessary for academic validation. This gap is particularly concerning given that industry practitioners increasingly report the need for flexible, auditable pipeline architectures that can adapt to changing regulatory requirements while maintaining operational efficiency.

\subsection{Industry–Academia Gap and the Need for Enhanced Patterns}

Contemporary literature reveals a persistent disconnect between academic research and industry practice in data pipeline engineering. Academic work often concentrates on theoretical optimizations or narrow performance metrics, while industry reports and technical blogs emphasize implementation details and tooling \cite{munappy2020challenges, liquibase2024governance}. As a result, practitioners have access to abundant "how-to" content but lack systematic guidance on applying reusable best practices or viewing pipelines through the lens of formal design patterns.  

For classical ETL and ELT processes, a mature body of best practices and governance frameworks provides reliable reference models for implementation \cite{acceldata2024multi}. By contrast, enhanced patterns such as ETLT and ELTL remain underdeveloped: despite their growing importance in enterprise-scale deployments, they are rarely discussed in terms of versioning, data contracts, continuous quality monitoring, or reproducibility. This gap leaves organizations vulnerable to governance failures and compliance risks, as such concerns are often treated as afterthoughts rather than first-class design elements. Additionally, the gap leaves data engineers without clear guidance, forcing them to rely on ad hoc scripts and point solutions that are difficult to maintain, scale, or audit. 

The urgency of enhanced patterns is evident in multi-cloud contexts, where traditional approaches struggle with consistency, governance, and lineage tracking across heterogeneous systems \cite{pace2019lineage, karanam2022lineage}. Surveys further highlight that existing academic frameworks inadequately address evolving schemas, the integration of batch and streaming paradigms, and the operational need for advanced observability and automated error detection \cite{schelter2018deequa, ballew2023auto}.  

These challenges highlight the need for fully defined design patterns that treat data quality contracts, versioning, lineage, and observability as built-in architecture elements. Simply tweaking ETL or ELT techniques won’t suffice, because it only improves existing methods rather than introducing new, reusable patterns. What we need are formally defined patterns that integrate data quality checks with flexible transformation steps, while also providing auditability, governance, and a boost to developer productivity.

To facilitate clarity and ease of reference, we denote the newly proposed design patterns with the suffix “++.” This notation highlights their evolutionary nature as extensions of existing paradigms while allowing for consistent identification throughout the paper. Accordingly, the ETLT++ and ELTL++ patterns proposed in this study in section \ref{sec:etltpp} and \ref{sec:eltlpp}, aim to fill this gap. Building on established principles of software design patterns, they provide reusable, auditable, and scalable solutions to recurring challenges in data pipeline engineering. By formalizing and extending enhanced pipeline patterns, this work bridges the divide between theoretical research and the operational realities of industry, equipping data engineering teams with systematic frameworks for trustworthy and efficient pipeline development.

\section{ETLT and ELTL Design Patterns}\label{sec:EtlElt}

Many works in the literature addressed challenges to existing ETL and ELT design pattern, which are recurrent problems that need to be solved. For instance, Hellerstein et al. \cite{hellerstein2017data} describe data wrangling as a ``fundamental bottleneck'' in analytics. Abadi et al. \cite{abadi2016modern} highlight that integration pipelines must balance quality, performance, and cost, often under cloud-scale constraints. Kleppmann \cite{kleppmann2017designing} emphasizes that data systems must be reliable, traceable, and reproducible to support long-lived applications. Karim et al. \cite{karim2018streaming} further show that with IoT and streaming contexts, heterogeneous and evolving schemas introduce constant stress on ingestion pipelines.  

Data engineers decide whether to transform data before or after loading based on things like how clean the source data is, whether they must keep the raw data for audits, how fast the pipeline needs to run, the compute resources available, and any governance or tracking requirements.

Based on the challenges aforementioned, it's quite evident from both literature and practice that a common picture emerges: ETL and ELT patterns, while foundational, are often insufficient. They do not fully address the conflicting requirements of modern platforms:
\begin{itemize}
  \item Ensuring data quality and preserving raw fidelity.  
  \item Balancing compliance (lineage, auditability) with performance (fast consumer access).  
  \item Supporting reprocessing at scale while containing cost growth from raw storage.  
  \item Allowing multiple teams to share pipelines without creating duplication or drift.  
\end{itemize}

Based on our practical experience in designing data pipelines, developing large-scale data platforms (e.g., for sustainability reporting and financial compliance), we repeatedly encountered these tensions.  In some cases, early transformation was necessary to stop ``bad data'' at the edge. In others, strict audit requirements mandated retention of the unaltered raw feed. 
Neither ETL nor ELT alone provided a generalizable, reusable approach to these recurring issues.  

Although data engineers routinely mix and match these approaches to handle real-world ingestion needs, the hybrid workflows known as ETLT (Extract, Transform, Load, Transform) and ELTL (Extract, Load, Transform, Load) have never been formally defined as design patterns. This motivates our formalization of two hybrid patterns: 
\textbf{ETLT} (Extract, Transform, Load, Transform) and \textbf{ELTL} (Extract, Load, Transform, Load).  
We declare and frame them in this paper as \emph{design patterns} because they describe not just technical steps, but the problem context, trade-offs, and consequences of adoption.  

In the next sub-sections formally define ETLT and ELTL as data ingestion design patterns, which are a reusable, high-level solution for how data is collected, imported, and moved from various sources into a system and we lay the foundation for their enhanced forms (ETLT++ and ELTL++), defined in section \ref{sec:etltpp} and \ref{sec:eltlpp}, which address further gaps in reproducibility, governance, and usability.

\subsection{ETLT Pattern (Extract, Transform, Load, Transform)} \label{sec:etlt_standard}

\textbf{Problem context} In many integration scenarios, the primary difficulty lies in the quality of incoming data. When data originates from multiple, heterogeneous sources—such as legacy transactional databases, external APIs, and semi-structured files—it is common to encounter inconsistencies, missing values, duplicate records, or schema mismatches. If such data is ingested ``as is,'' downstream transformations may produce unreliable results, business rules may fail, and debugging becomes both time-consuming and costly. In large-scale compliance or financial systems, even minor quality issues at ingestion can cascade into significant operational risks. This is a well-recognized challenge in the data engineering literature, where early cleansing and validation have been identified as critical for ensuring trust in downstream pipelines \cite{hellerstein2017data, abadi2016modern}. Grounded in our hands-on experience as data engineering practitioners, designing data platforms for sustainability reporting, a lack of early validation meant that entire reporting cycles were delayed because faulty source records had already polluted the central repository. 

\textbf{Definition.} The ETLT pattern represents a multi-stage integration paradigm that decouples data quality operations from business-specific transformations. Data is first extracted from heterogeneous sources such as databases, APIs, and flat files. An initial transformation stage, denoted \(T_1\), applies cleansing, validation, and normalization to enforce consistency and data contracts. Only after this quality gate does the data move to the \emph{Load} stage, where it is persisted in a structured repository (e.g., data lake or warehouse). A second transformation stage, \(T_2\), then applies business rules, enrichments, and schema shaping to prepare the data for analytical consumption. 

By isolating quality concerns in \(T_1\), ETLT ensures that downstream business transformations are not contaminated by invalid inputs. This separation also enables:
\begin{itemize}
  \item Parallel execution of business logic once high-quality intermediates are available.
  \item Easier debugging and fault isolation, since quality failures are detected before load.
  \item Deterministic replay of \(T_2\) without re-extracting or re-cleaning source data.
\end{itemize}

\begin{center}
\begin{tikzpicture}[node distance=0.8cm, >=latex']

\node (extract) [rectangle, draw, minimum width=1.2cm, minimum height=1cm] {\footnotesize Extract (Sources)};
\node (transform1) [rectangle, draw, right=of extract, minimum width=1cm, minimum height=1cm] {\footnotesize Transform$_1$ (Quality)};
\node (load) [rectangle, draw, right=of transform1, minimum width=1.4cm, minimum height=1cm] {\footnotesize Load};
\node (transform2) [rectangle, draw, right=of load, minimum width=1cm, minimum height=1cm] {\footnotesize Transform$_2$ (Business)};

\draw[->, thick] (extract) -- (transform1);
\draw[->, thick] (transform1) -- (load);
\draw[->, thick] (load) -- (transform2);

\end{tikzpicture}
\end{center}

ETLT is especially appropriate in settings where data quality cannot be taken for granted. In environments that combine feeds from diverse upstream systems, or where source reliability is low, the pattern ensures that invalid or non-compliant records are detected before they are allowed into the central repository. This makes ETLT valuable in regulatory contexts where data contracts are enforced at the point of entry, such as financial or healthcare reporting systems. It also proves effective in real-time scenarios, where heterogeneous batch and streaming data must be reconciled consistently before use. Finally, for large-scale migration projects—where historical data from multiple inconsistent sources is consolidated—ETLT offers a disciplined method for staging and validating inputs before long-term storage. In short, ETLT provides a structured and reusable solution whenever the main organizational risk stems from low-quality or inconsistent input data.  This formalization highlights ETLT as a reusable design pattern, not only an operational choice, that systematically enforces quality before persistence and business use.

\subsection{ELTL Pattern (Extract, Load, Transform, Load)}

\textbf{Problem context} In data engineering contexts, the primary concern is not input quality of data, but data preservation and flexibility. Data engineers in finance, healthcare, and IoT domains often face strict requirements to retain every record exactly as it was received, even if flawed, for purposes of compliance, lineage, or forensic analysis. This demand stems from the need to re-run historical computations, to audit past states of the data, or to reconstruct events for regulatory authorities. In fast-evolving domains such as IoT and streaming systems, schemas may change unpredictably, and transformations applied too early may discard valuable information that later becomes relevant \cite{kleppmann2017designing, karim2018streaming}.  In our work building data platforms for sustainability and energy monitoring, we repeatedly encountered requirements to regenerate reports using the exact raw data from months or years earlier. Without a faithfully preserved raw zone, such requests would have been impossible to satisfy. 

\textbf{Definition} The ELTL pattern is a dual-loading architecture that emphasizes \emph{preservation of raw data alongside performance-optimized outputs}. Extraction is performed with minimal preprocessing to retain fidelity. The data is then loaded into a \emph{raw zone} (e.g., a landing area in a lakehouse), creating immutable records. A comprehensive transformation stage follows, leveraging cloud-scale compute for enrichment, joining, and aggregation. The resulting curated data is then loaded again into optimized structures (e.g., partitioned or columnar tables) for consumption.

ELTL is motivated by:
\begin{itemize}
  \item \emph{Lineage and compliance}: retaining raw data and its schema / structure guarantees a full audit trail.
  \item \emph{Flexibility}: schema evolution is easier since raw data remains intact.
  \item \emph{Multi-team reuse}: different consumers can apply distinct transformations starting from the raw layer.
  \item \emph{Cost optimization}: storage is decoupled from compute, allowing selective materialization.
\end{itemize}

\begin{center}
\begin{tikzpicture}[node distance=0.8cm, >=latex']

\node (extract) [rectangle, draw, minimum width=1.2cm, minimum height=1cm, align=center] {\footnotesize Extract (Sources)};
\node (load1) [rectangle, draw, right=of extract, minimum width=1.2cm, minimum height=1cm, align=center] {\footnotesize Load$_1$ Raw data};
\node (transform) [rectangle, draw, right=of load1, minimum width=1.6cm, minimum height=1cm, align=center] {\footnotesize Transform};
\node (load2) [rectangle, draw, right=of transform, minimum width=1.4cm, minimum height=1cm, align=center] {\footnotesize Load$_2$ Optimized data};

\draw[->, thick] (extract) -- (load1);
\draw[->, thick] (load1) -- (transform);
\draw[->, thick] (transform) -- (load2);

\end{tikzpicture}
\end{center}

ELTL is particularly well suited to environments where data must be preserved in its original form before any processing, because it guarantees complete auditability, reproducibility, and cross‐team reuse. By loading raw data first, ELTL creates an immutable archive that answers questions like “What did the system record on August 1st?” without relying on reconstructed or overwritten records. In IoT and streaming contexts, where sensor feeds evolve quickly and future use cases are hard to anticipate, preserving raw data ensures that nothing is prematurely discarded.  Moreover, in multi-tenant data platforms, where different teams or departments apply distinct transformations to the same source data, the raw zone serves as a shared foundation, reducing duplication and drift. Thus, ELTL provides a structured and reusable solution for organizations whose main risk lies not in bad input data, but in losing fidelity, lineage, or flexibility for future reprocessing.  

\subsection{Summary of ETLT and ELTL design patterns}

ETLT and ELTL patterns can be formally considered design patterns because they represent generalized, reusable solutions to recurring challenges in data integration, quality assurance, and analytical optimization. 

ETLT stops low quality data before it lands, simplifying governance and reducing downstream confusion, making it ideal when upstream contracts exist. ELTL, by contrast, always preserves what arrived, enabling maximum lineage and rollback, fitting compliance-heavy and exploratory settings. Table~\ref{tab:patterns} provides a structured comparison.

\begin{table}[ht]
\centering
\caption{When to Choose ETLT vs.\ ELTL}
\label{tab:patterns}
\begin{tabular}{l p{4.5cm} p{4.5cm}}
\toprule
\textbf{Aspect} & \textbf{ETLT (Validate First)} & \textbf{ELTL (Load First)} \\
\midrule
Data Quality Control &  
Reject or quarantine bad records at ingestion, keeping only clean data. &  
Store everything, then clean or transform as needed later. \\[6pt]

Audit &  
Cannot replay rejected data, but downstream is always clean. &  
Full history of raw inputs—ideal for audits and time travel. \\[6pt]

Flexibility &  
Simplifies downstream logic by guaranteeing data quality early. &  
Allows different teams to apply custom transformations on the same raw data. \\[6pt]

Cost &  
Smaller storage needs for raw zone since bad data is filtered out. &  
Larger raw storage footprint but cheaper to keep raw data only once. \\[6pt]

Team Collaboration &  
Central team enforces contracts at the “gate.” &  
Multiple teams reuse a shared raw layer without re-ingesting data. \\[6pt]

Use Cases &  
High-quality data needed immediately (e.g., operational dashboards). &  
Compliance, forensic analysis, or exploratory research (e.g., finance, healthcare, IoT). \\
\bottomrule
\end{tabular}
\end{table}

ETLT is most effective when data quality issues are frequent and need to be addressed independently from business transformations, whereas ELTL is ideal when regulatory compliance, auditability, or multi-team analytical usage is required. Collectively, these patterns exemplify structured approaches to daily data engineering problems, offering scalability, robustness, and maintainability in complex, heterogeneous, and large-scale data environments. However, despite their strengths, both patterns often treat validation, lineage, and quality monitoring as afterthoughts rather than built-in features. They lack a unified mechanism for mandatory contract enforcement, deterministic replay, lineage capture, and measurable service-level objectives for data quality. In the next section (\ref{sec:etltpp}) we extends ETLT into ETLT++, adding explicit contract gating, deterministic replay, lineage capture, and quality SLOs to make pipelines reliably reproducible and audit-ready.

From a reader’s perspective, one might question whether ETLT is merely “ETL with an additional validation step” and whether ELTL is simply “ELT with staging.” This concern is understandable, given the superficial resemblance. However, the distinction lies in the explicit separation of validation as a first-class design concern and the structured use of staging to support traceability and reusability. These refinements elevate the patterns beyond minor variations, positioning them as generalizable and reusable solutions within data ingestion design. 

\section{ETLT++: Ensuring reliable and understandable Data Pipelines}
\label{sec:etltpp}

In this section, we propose \textbf{ETLT++}, an enhanced version of the ETLT Design Pattern that we define in Section \ref{sec:etlt_standard}. While ETLT separates quality checks from business logic, ETLT++ extends this by explicitly embedding contracts, checkpoints, rewindability, and quality monitoring into the pattern. The novelty lies in moving from ad-hoc hybrid workflows to a formally structured design pattern that guarantees reproducibility, auditability, and continuous quality assurance.

We define an ETLT++ Design Pattern as a sequence of connected stages:
\[
P = \langle E, C, T_1, L, T_2, O \rangle
\]
where each symbol represents:

\begin{itemize}
    \item \textbf{$E$ (Extract from Sources):} Raw ingredients—spreadsheets, databases, or real-time sensor feeds.
    \item \textbf{$C$ (Data Contract Loading):} Retrieve the JSON‐based data contract from a centralized registry. This contract specifies required fields, value ranges, formats, and rule severities (hard vs.\ soft).
    \item \textbf{$T_1$ (Validation and Cleaning):} Enforce the contract:  
      \begin{itemize}
        \item Record‐level checks quarantine any record violating a hard rule and log warnings for soft‐rule breaches.
        \item Batch‐level check halts ingestion if any hard violations occur.
      \end{itemize}
    \item \textbf{$L$ (Load into Versioned Raw Storage):} Store validated records in a lineage‐aware raw zone that preserves every version and timestamp.
    \item \textbf{$T_2$ (Business Logic and Transformation):} Operations that convert raw data into structured, analysis‐ready datasets, e.g., aggregations, enrichments, or historical change tracking, using reusable transformation templates.
    \item \textbf{$O$ (Outputs):} Publish curated datasets to downstream consumers such as dashboards, reports, or machine‐learning pipelines.
\end{itemize}

Unlike classic ETLT, ETLT++ defines design obligations: contracts must exist at ingress, validated data must be versioned in raw form, business logic must be replayable, and pipeline health must be continuously monitored. These obligations are what make ETLT++ a reusable \emph{design pattern} rather than 
a one-off implementation.

\begin{figure}[!ht] \centering \includegraphics[scale=0.23]{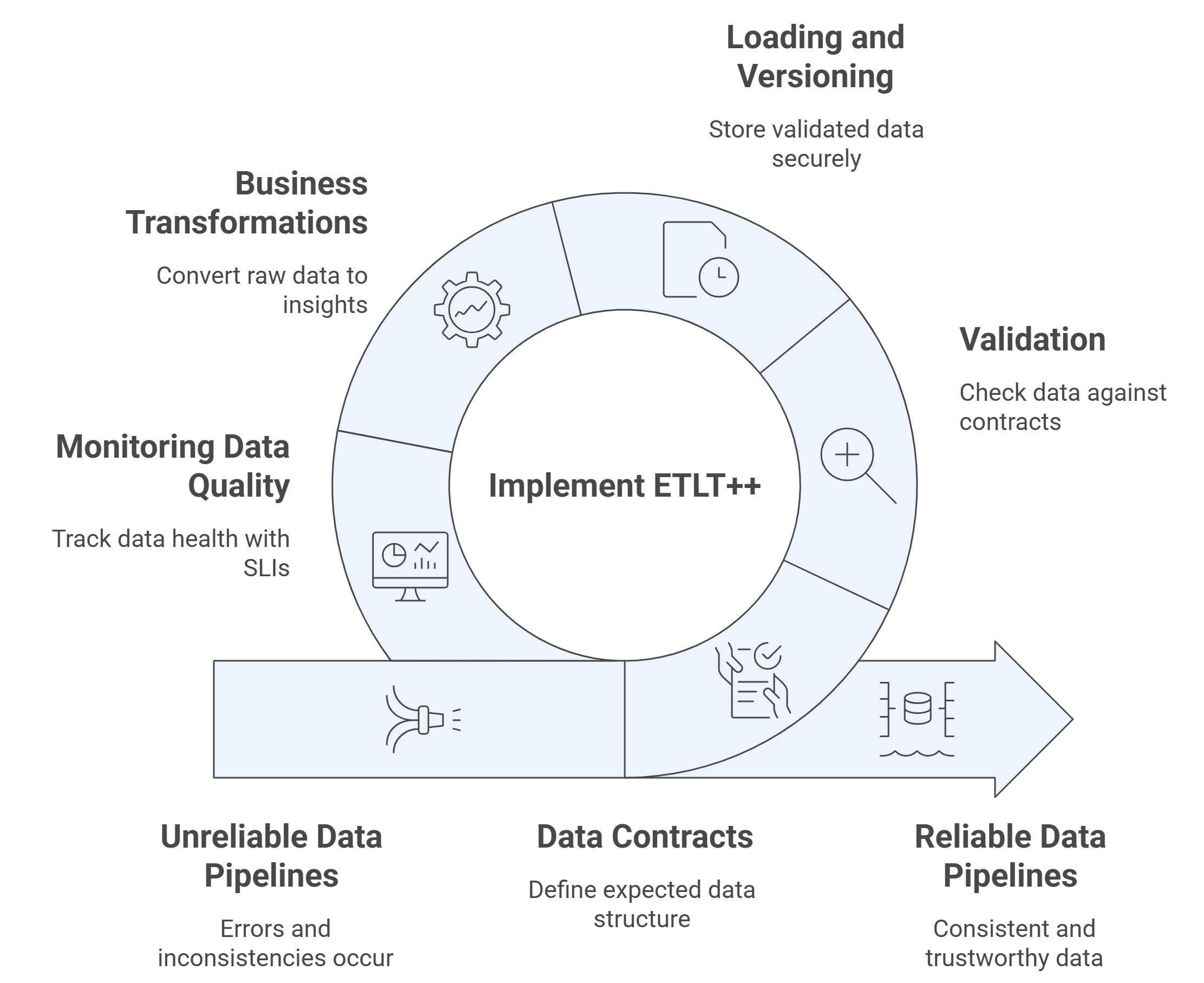} \caption{ETLT++: Steps for a reliable pipeline} \label{fig:etlt} \end{figure}

\subsection{Data Contracts}

In practice, one of the most critical problems in data engineering is that the raw data arriving from multiple, heterogeneous sources is often incomplete, inconsistent, or violates implicit assumptions. Without a mechanism to block such inputs, corrupted data silently enters the repository and contaminates downstream analytics and machine learning models. This is not only a technical issue but an organizational one: teams waste time and efforts debugging symptoms instead of addressing root causes. Drawing from our professional practice in data engineering, designing data pipelines for sustainability reporting, a single upstream error (e.g., a negative energy consumption value) propagated unchecked into financial dashboards, delaying regulatory submissions by weeks. Such examples illustrate why a more formal mechanism is needed.

Unlike ETLT, where validation mechanisms may be implicit or ad hoc, ETLT++ makes the use of data validations defined in data contracts as mandatory and an explicit safeguard against corrupted or non-conforming inputs. A data contract defines the structural, semantic, and quality rules that incoming data must satisfy to be considered valid \cite{schelter2018deequa}. While data contracts have been introduced in the literature as a general mechanism for data validation, in ETLT++ they function as binding agreements between producers and consumers that cannot be bypassed. In the following paragraphs, we provide a detailed description of data contracts and their role in ensuring reliability in data pipelines.

Data contracts are not novel, they are a well-established mechanism in data governance frameworks \cite{batini2009methodologies, atlan2025framework}. However, in existing systems they are often optional or implemented inconsistently. In ETLT++ and ELTL++, we make data contracts a mandatory, first-class design element. A data contract is a static specification of rules that every dataset must satisfy before entering the pipeline. We store contracts as JSON documents in a centralized metadata registry (e.g., a Git-backed catalog). A contract for a customer transactions dataset might include:

\begin{itemize}
    \item Required fields: Certain columns or attributes must exist in every dataset. For example, in a customer transactions dataset, the fields \texttt{customer\_id}, \texttt{transaction\_date}, and \texttt{transaction\_amount} might be mandatory. Missing fields are considered violations of the contract.
    
    \item Value ranges and types: Numeric fields must lie within reasonable limits (e.g., \texttt{age} $\geq 0$, \texttt{transaction\_amount} $\geq 0$). String fields may need to respect specific formats, such as email addresses or phone numbers. These rules prevent nonsensical or malformed data from entering the system.
    
    \item Rule severity: Rules can be classified as \emph{hard} or \emph{soft}. Hard rules are strict constraints: if violated, the data should not enter the pipeline. Soft rules are advisory: violations trigger warnings but do not block processing. This distinction allows flexibility while maintaining data safety.
\end{itemize}

\subsection{Validation: Enforcing Contracts}

We introduce a mandatory validation stage that enforces data contracts before any loading occurs. Validation operates at both record and batch granularity:
\begin{enumerate}
  \item Retrieve contract: Load the JSON contract from the metadata registry.
  \item Record‐level evaluation: For each record \(r\) in the incoming batch:
    \begin{itemize}
      \item Compute a Boolean indicator \(v_{i,r}\in\{0,1\}\) for each hard rule \(i\):  
        \(v_{i,r}=1\) if record \(r\) violates rule \(i\), else 0.
      \item If any \(v_{i,r}=1\), mark record \(r\) as \emph{quarantined} and skip further processing for that record.
      \item If a soft rule is violated, log a warning but allow the processing to continue.
    \end{itemize}
  \item Batch‐level decision:
    Compute total hard violations:  
    \[
      V = \sum_{r}\sum_{i}v_{i,r}.
    \]  
    If \(V>0\), halt ingestion of the entire batch and flag it for correction; otherwise, proceed to loading.
\end{enumerate}

\textbf{Example:} Suppose we receive a batch of five customer records:
\begin{center}
\small
\begin{tabular}{p{1.4cm} p{1.3cm} p{2.5cm} c p{3cm}}
\toprule
\textbf{client\_id} & \textbf{amount} & \textbf{email} & \textbf{Status} & \textbf{Reason} \\
\midrule
1001 & 50  & alice@example.com & Pass       & Validated successfully. \\
1002 & -20 & bob@example.com   & Quarantined & Hard rule violated: negative amount. \\
1003 & 30  & (missing)         & Pass       & Soft rule violated: missing email. \\
1004 & 0   & carol@example.com & Pass       & Validated successfully. \\
1005 & 10  & dave@example.com  & Pass       & Validated successfully. \\
\bottomrule
\end{tabular}
\end{center}

\begin{itemize}
    \item The second row violates the hard rule on \texttt{transaction\_amount} (negative value).  
    \item The third row violates the soft rule on \texttt{customer\_email} (missing email).  
\end{itemize}

Validation results:
\begin{itemize}
    \item Hard rule violation triggers quarantine of the batch. No further processing occurs until the issue is corrected.  
    \item Soft rule violation generates a warning in logs for later inspection, but does not block processing.  
\end{itemize}

Only record 1002 violates a hard rule and is quarantined. Since \(V=1\), the batch ingestion is halted until the issue is resolved. Soft violations (record 1003) generate warnings but do not block processing.

The effect of this contract-validation combination is that pipelines become predictable: invalid records are stopped at the edge, warnings are logged for follow-up, and only trustworthy data reaches the repository. In other words, in ETLT++, data quality is not an optional feature but a mandatory property of a robust modern data platform.

\subsection{Loading and Versioning ($L$)}

In many data pipelines, the loading phase is treated as a mere ``black box'' operation in which data is inserted into a database or lake without further design considerations. This oversimplification creates two recurring problems. First, many widely used storage systems do not support versioning natively, which means that once data is overwritten, prior states are lost permanently. This makes it impossible to reconstruct what the dataset looked like at a given point in time, undermining reproducibility and compliance. Second, even when modern table formats such as Delta Lake, Apache Iceberg, or Hudi are available, versioning features are often poorly configured or entirely neglected. In both cases, the consequence is the same: teams are unable to reproduce historical analyses, perform reliable audits, rewind business flows, or debug transformations against the data state that originally triggered the issue. In our own professional practice, we have seen dashboards produce different values for the same query depending on when it was run, simply because versioning was not enforced. Such inconsistencies erode trust in the data platform and increase operational risk.

As a result, reproducibility and traceability are compromised, making it impossible to answer questions such as: \textit{“What did the data look like last week when the report was generated?”} or\textit{ “Which records changed since last audit?”}. This gap in ETLT has direct consequences for regulatory compliance and debugging.  

ETLT++ therefore treats \textbf{versioned, append-only loading} as mandatory. Data is persisted immutably: once written, records are never deleted or modified but only appended. This guarantees a full lineage of states over time. 

Modern table formats such as \textbf{Delta Lake} and \textbf{Apache Iceberg} natively support these capabilities, including time-travel queries, snapshot isolation, and transactional consistency \cite{delta_lake_time_travel, iceberg_spec}. 

We provide in our ELTL++ design patterns practical guidance for systems that lack native versioning capabilities. For systems that do not support versioning control over data, best practices can emulate it: in relational databases, for example, versioning can be emulated by introducing \texttt{valid\_from} and \texttt{valid\_to} fields. Each update closes the validity of the current record by setting the \texttt{valid\_to} timestamp and inserts a new row with an updated validity interval. In object stores, immutability can be achieved by enforcing file-level partitioning by load date (e.g., storing each batch with a timestamp like \texttt{/transactions/2025-08-27/}). Even in cases where storage costs or performance considerations prevent full lineage retention, ETLT++ encourages partial versioning strategies not for the whole data, but for some specific data entities that are considered to be important to be versioned, such as maintaining detailed history only for critical financial columns or limiting full lineage to a specific retention window (e.g., the last two years), while archiving older snapshots.  

By embedding data versioning control practices into the design pattern, ETLT++ ensures that versioning is always present, whether natively supported or emulated. This eliminates the fragility of ad-hoc approaches and guarantees that analysts, auditors, and engineers can always “time-travel” through the dataset to reproduce, validate, and trust their results.

\emph{Example} 
Consider a repository $R$ that stores daily customer transactions with versioning enabled. Each record includes two metadata fields: \texttt{valid\_from} (the load timestamp) 
and \texttt{valid\_to} (set to NULL for the active version).  

Suppose yesterday’s storage $R$ contains:

\begin{center}
\begin{tabular}{l|l|l|l|l}
\textbf{transaction\_id} & \textbf{customer\_id} & \textbf{amount} & \textbf{valid\_from} & \textbf{valid\_to} \\
\hline
T001 & 1001 & 50 & 2025-08-26 & NULL \\
T002 & 1002 & 20 & 2025-08-26 & NULL \\
\end{tabular}
\end{center}

Today we receive a new batch $V$ with two records:

\begin{center}
\begin{tabular}{l|l|l}
\textbf{transaction\_id} & \textbf{customer\_id} & \textbf{amount} \\
\hline
T002 & 1002 & 20 \\
T003 & 1003 & 30 \\
\end{tabular}
\end{center}

The versioning rules are applied as follows:

\begin{itemize}
    \item Record \texttt{T002} already exists with the same values. 
    Its version is preserved, no new row is added.  
    \item Record \texttt{T003} is new. 
    It is inserted into $R$ with \texttt{valid\_from = 2025-08-27} and \texttt{valid\_to = NULL}.
\end{itemize}

The repository $R$ now contains:

\begin{center}
\begin{tabular}{l|l|l|l|l}
\textbf{transaction\_id} & \textbf{customer\_id} & \textbf{amount} & \textbf{valid\_from} & \textbf{valid\_to} \\
\hline
T001 & 1001 & 50 & 2025-08-26 & NULL \\
T002 & 1002 & 20 & 2025-08-26 & NULL \\
T003 & 1003 & 30 & 2025-08-27 & NULL \\
\end{tabular}
\end{center}

If tomorrow the amount for \texttt{T002} changes to 25, the system does not overwrite the old record. Instead:

\begin{center}
\begin{tabular}{l|l|l|l|l}
\textbf{transaction\_id} & \textbf{customer\_id} & \textbf{amount} & \textbf{valid\_from} & \textbf{valid\_to} \\
\hline
T001 & 1001 & 50 & 2025-08-26 & NULL \\
T002 & 1002 & 20 & 2025-08-26 & 2025-08-28 \\
T002 & 1002 & 25 & 2025-08-28 & NULL \\
T003 & 1003 & 30 & 2025-08-27 & NULL \\
\end{tabular}
\end{center}
\begin{itemize}
    \item The existing row for \texttt{T002} is closed by setting \texttt{valid\_to = 2025-08-28}.  
    \item A new row is inserted with \texttt{amount = 25}, \texttt{valid\_from = 2025-08-28}, and \texttt{valid\_to = NULL}.
\end{itemize}

This creates two explicit versions of \texttt{T002}, one valid until 2025-08-28 and one active afterwards. \emph{Why is this useful?} Because analysts can now issue queries such as:
\begin{verbatim}
-- State of data as of August 26
SELECT * FROM transactions WHERE valid_from <= '2025-08-26' 
  AND (valid_to IS NULL OR valid_to > '2025-08-26');
\end{verbatim}

This returns only the records valid on that date, faithfully reconstructing the dataset as it was originally ingested.  

Through this mechanism, every state of the data can be recreated at any historical point. This enables reproducibility of past analyses, compliance with audit requests, and debugging of transformations without risk of contamination from newer changes. In short, versioning transforms loading into a transparent, auditable, and rewindable process.

ETLT++ transforms ETLT into a robust design pattern with the following contributions:
\begin{itemize}
  \item Contracts as obligations: enforceable agreements, not optional schemas.
  \item Versioned raw storage: immutability, traceability, and reproducibility by design.
  \item Rewindable business logic: deterministic transformations that can be reapplied.
  \item Continuous monitoring: SLIs and SLOs integrated into the pipeline fabric.
\end{itemize}

Together, these properties ensure that ETLT++ pipelines are reliable, transparent, and auditable, a novel contribution beyond existing ETL/ELT practices.

\section{ELTL++: Raw-Preserving Pipelines for Modern Data Platforms} \label{sec:eltlpp}

The ELTL pattern (Extract, Load, Transform, Load) has gained wide adoption in large-scale data engineering because it preserves raw fidelity: data is loaded into a storage layer before undergoing business transformations. 
This ordering enables reprocessing from first principles, ensures compliance with audit requirements, and offers flexibility for multiple downstream consumers.  

However, while it is attractive in theory, practitioners have identified recurring 
shortcomings when applying ELTL in modern platforms:
\begin{itemize}
  \item \textbf{Raw data sprawl.} The first loading stage ($L_1$) tends to grow 
        without limit, leading to so-called ``data swamps'' where useful information becomes difficult to find and manage.
  \item \textbf{Inconsistent transformations.} Different data engineering teams often reimplement similar transformation logic, producing redundant or contradictory curated datasets. Significant effort is then required to reconcile these discrepancies and restore a single source of truth.
  \item \textbf{Delayed usability.} Because transformations are deferred until after $L_1$, data consumers may face long wait times, and additional delays caused by resolving contradictory outputs, before reliable, curated datasets become available for analysis.
  \item \textbf{Weak governance.} Without systematic lineage, metadata, or contracts, 
        it is difficult to establish trust in curated outputs or reproduce past results.
\end{itemize}

These challenges illustrate that ELTL in its baseline form is not sufficient to support the demands of modern, multi-tenant, cloud-native data platforms.
We therefore propose ELTL++, an enhanced design pattern that addresses these gaps through four structured enhancements: (i) smart raw data management, (ii) standardized and versioned transformations, (iii) dual loading with a curated semantic layer, and (iv) embedded governance and observability.  

\begin{figure}[!ht] \centering \includegraphics[scale=0.23]{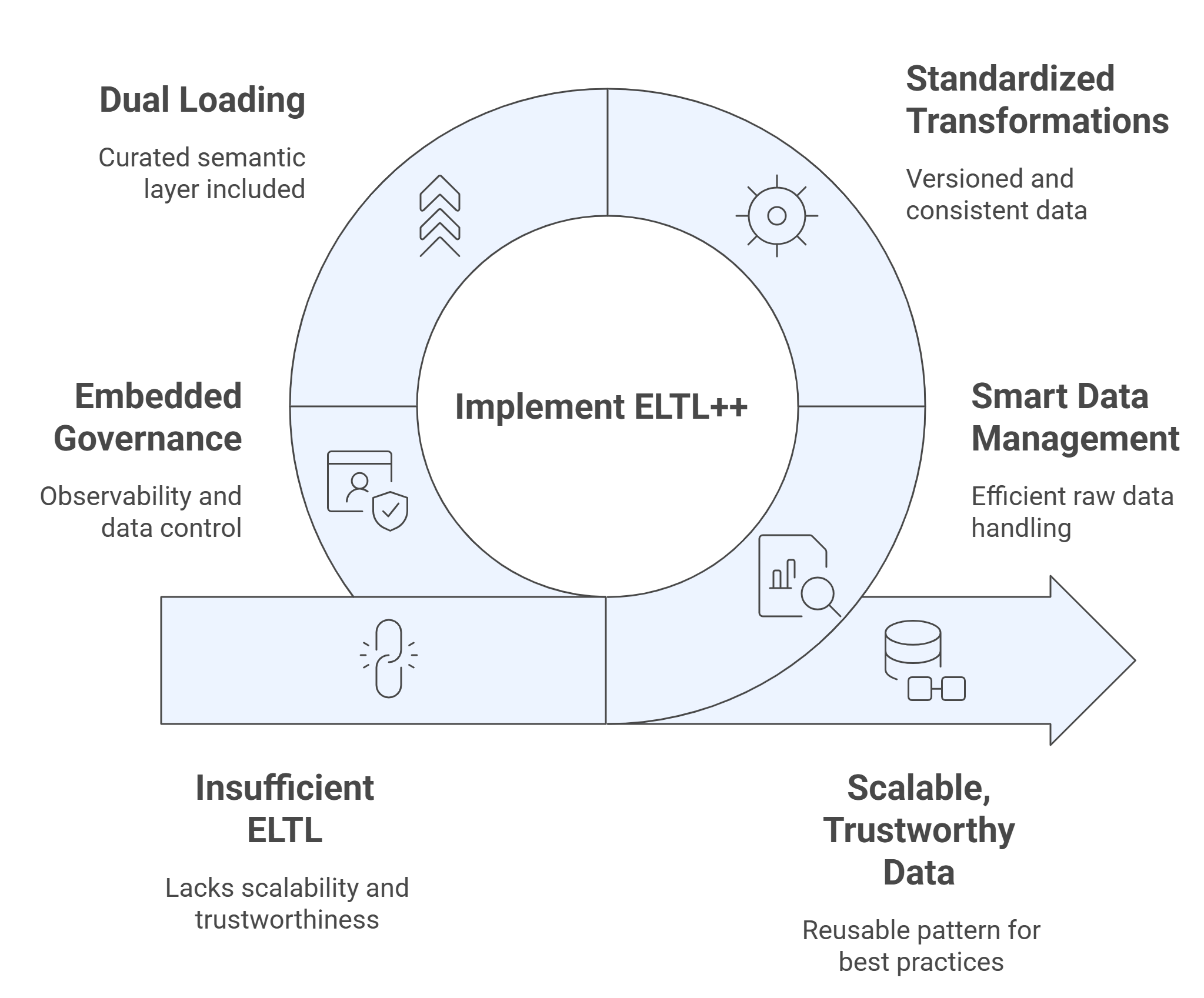} \caption{ELTL++: Steps for a reliable pipeline} \label{fig:eltl} \end{figure}

The purpose of ELTL++ is not to replace ELTL but to provide a reusable, systematic pattern that captures best practices for \emph{scalability, trustworthiness, and consumer usability}.  

\subsection{Raw Layer Management (L1)}

\textbf{Problem}  
In baseline ELTL, the first load ($L_1$) is simply a raw dump of everything received.  
While this ensures fidelity, it creates a common challenge known as the \emph{data swamp}. Without retention policies or indexing, raw data grows without bound, becomes expensive to store, and is practically impossible to navigate. This is especially problematic in IoT or streaming environments, where billions of events are produced daily.  

ELTL++ redefines $L_1$ as a managed raw layer. Instead of a blind archive, the raw zone is equipped with policies and metadata to make it navigable, sustainable, and secure. Its key properties include:
\begin{itemize}
  \item \emph{Metadata-driven ingestion} Usage of a parametrized pipeline, built on a metadata table, as defined in a formalized Design Pattern. \cite{Rucco2025}
  \item \emph{Retention and tiering} Policies define how long raw data remains in fast storage 
        (e.g., 90 days) before being moved to colder, cheaper storage. 
        This reduces cost without sacrificing reproducibility.
  \item \emph{Access policies} Sensitive data is automatically tagged and masked, 
        preventing accidental misuse while still retaining original records for audits.
\end{itemize}

\emph{Example} 
Consider a utility company that inhates 5TB (5120GB) of IoT sensor data per month into a raw zone under a classical ELTL pipeline. Using Azure Blob Storage pay-as-you-go rates—\$0.021 per GB-month for Hot tier and \$0.00099 per GB-month for Cool tier \cite{azureblobpricing}, the costs are:

\begin{itemize}
  \item ELTL (all data in Hot tier):
    \begin{itemize}
      \item \emph{Hot data:} The most recent or frequently accessed subset stored in high-performance, low-latency Hot tier (e.g., operational dashboards).
      \item Storage: $5120\text{GB}\times\$0.021/\text{GB-month} = \$107.52/\text{month}\approx\$3.58/\text{day}$
      \item Compute and operations overhead: \$5.00/day
      \item \emph{Total cost:} \$8.58/day
    \end{itemize}
  \item ELTL++ (10\% Hot, 90\% Cool):
    \begin{itemize}
      \item \emph{Hot data:} 512GB of recent data in Hot tier for fast, interactive queries.
      \item \emph{Cool data:} 4608GB of older or infrequently accessed data in Cool tier (lower-cost, slightly higher-latency storage for archival and compliance).
      \item Hot storage: $512\text{GB}\times\$0.021/\text{GB-month} = \$10.75/\text{month}\approx\$0.36/\text{day}$
      \item Cool storage: $4608\text{GB}\times\$0.00099/\text{GB-month} = \$4.56/\text{month}\approx\$0.15/\text{day}$
      \item Reduced compute and operations (smaller hot zone): \$0.10/day
      \item \emph{Total cost:} \$0.61/day
    \end{itemize}
\end{itemize}

\begin{table}[ht]
\centering
\caption{Cost Comparison: ELTL vs.\ ELTL++}
\begin{tabular}{|p{1.5cm}|p{1.5cm}|p{1.5cm}|p{1.5cm}|p{1.5cm}|p{1.5cm}|}
\toprule
Pipeline & Hot Data & Cool Data & Storage Cost/day & Compute /day & Total/day \\
\midrule
ELTL      & 5120GB & 0GB    & \$3.58          & \$5.00         & \$8.58   \\
ELTL++    &  512GB & 4608GB & \$0.36+\$0.15   & \$0.10         & \$0.61   \\
\bottomrule
\end{tabular}
\end{table}

By indexing metadata and applying an archival policy, ELTL++ reduces hot‐storage by 90\% and cuts total daily cost by over 90\%, while preserving older data affordably in Cool storage for compliance and historical analysis.

\subsection{Dual Loading with a Curated Semantic Layer (L2)}

\textbf{Problem}  
A major limitation of baseline ELTL pipelines is that business users often face two equally undesirable options:  \begin{itemize}
    \item Query raw data directly from $L_1$, which is complex, inconsistent, and unsuitable for non-technical analysts; or  
 \item  Wait for delayed and ad-hoc curated datasets that are manually produced by engineering teams.

\end{itemize}

In both cases, the promise of self-service analytics and rapid business decision-making is undermined.  

Raw data, while essential for fidelity and auditability, is not user-friendly. It typically contains nested schemas, system-generated identifiers, or event logs that require significant technical knowledge to interpret. Business stakeholders, however, expect semantic constructs: ready-to-use metrics, dimensional models, and standardized fact tables. Without this, organizations experience slow reporting cycles, inconsistent definitions of key indicators (e.g., revenue, churn, active users), and heavy reliance on scarce engineering resources.  

ELTL++ introduces a second loading stage, $L_2$, explicitly defined as a \emph{curated semantic layer} to make raw data inquiring easier. In $L_2$ step, raw data from $L_1$ is systematically transformed and published into consumer-ready datasets. The semantic layer is designed according to the following principles: 

\begin{itemize}
  \item \textbf{Dimensional modeling:} Data in $L_2$ is organized into fact and dimension tables following Kimball’s star and snowflake schemas \cite{kimball2002data}. This structure makes it intuitive to slice and aggregate data across business dimensions, for example, computing sales totals by region or customer counts by segment.
  \item \textbf{Metric standardization:} the role of the semantic layer is to link between the user-defined metrics and an sql query that retrieves data from the previous layer (database, or any other storage method defined in $L_1$). itself defines and stores pre-computed KPIs such as ``monthly recurring revenue'' or ``active users'', according to a single, organization-wide definition. In this way, all downstream dashboards and reports reference the same metric implementations, eliminating discrepancies.
  \item \textbf{Enabling self services:} $L_2$ exposes simplified, user-friendly schemas in which column names, hierarchies, and relationships are already aligned with business vocabulary. Analysts and BI tools can query data without deep technical knowledge of $L_1$.  
  \item \textbf{Reproducibility:} curated datasets are versioned and linked back to their sources in $L_1$, so that any semantic result can be traced and reproduced exactly.  
\end{itemize}

This dual loading strategies that we add in ELTL++ preserves the benefits of raw data retention at $L_1$ (fidelity, auditability, reprocessing), while delivering immediate usability at $L_2$ (business-friendly, standardized, and self-service ready).  

\emph{Example}  
Consider a retail platform ingesting daily transactions.  
In baseline ELTL, analysts must either query raw transaction logs or request ad-hoc transformations from engineers.  
In ELTL++, the pipeline operates as follows:  

\begin{verbatim}
-- Stage L1: immutable raw logs
INSERT INTO S1.transactions_raw
VALUES (txn_id, customer_id, product_id, amount, timestamp);

-- Stage L2: curated semantic layer
INSERT INTO S2.daily_revenue_by_region
SELECT region, SUM(amount) as revenue
FROM L1.transactions_raw tr
JOIN L1.customers_raw c ON tr.customer_id = c.id
GROUP BY region, DATE(timestamp);
\end{verbatim}

The result is a table \texttt{daily\_revenue\_by\_region} in $L_2$, which is refreshed automatically each night.  
From a business user’s perspective, this dataset already contains the relevant dimensions (region, date) and aggregated measures (revenue), without requiring them to parse raw transactions or join multiple tables (thanks to the semantic layer). 

Thus, the curated semantic layer in ETLT++ functions as the mediator between technical pipelines and business analytics. Engineers ensure fidelity and governance in $L_1$, while $L_2$ guarantees usability, standardization, and alignment with business language.  
This shift transforms ELTL pipelines from raw archives into platforms that directly support self-service BI, data democratization, and consistent decision-making across the organization.  

\section{Monitoring and Ensuring Data Quality}

In modern data pipelines, data quality is not optional, it is essential. Even with rigorous contracts, validation, and standardized transformations, pipelines can produce errors, delays, or inconsistencies due to upstream failures, missing data, schema changes, or human errors. Poor-quality data propagates downstream, compromising dashboards, reports, and machine learning models, and ultimately undermining business decisions. Therefore, ETLT++ and ELTL++ integrate continuous data quality monitoring as a fundamental step in the design pattern, ensuring that all derived datasets are trustworthy, complete, and timely.

Data quality challenges are pervasive. Raw datasets may arrive late or partially, contain invalid or inconsistent values, or violate agreed-upon schemas. These issues can disrupt downstream processes, leading to incorrect metrics, failed aggregations, or misleading analyses. Simply relying on manual checks or ad hoc validation is insufficient for modern, large-scale data platforms, where pipelines must operate reliably and autonomously.

To address these challenges, ETLT++ proposes a structured, reusable design pattern for monitoring and enforcing data quality. The first step is to define \emph{Service Level Indicators} (SLIs) that capture the main aspects of data quality. These dimensions—including freshness\cite{wang1996beyond}, completeness \cite{english1999improving}, accuracy \cite{pipino2002data}, and contract adherence \cite{dehghani2022data}—are well-established in data quality literature and frameworks \cite{batini2009methodologies}. However, our key contribution lies in making these quality assessments \emph{mandatory} rather than optional within the pattern design. Unlike traditional approaches where data quality monitoring is often treated as an add-on or afterthought, ETLT++ embeds these SLIs as required architectural components that must be implemented for the pattern to be considered complete.

\begin{itemize} \item \textbf{Freshness:} This metric measures how up-to-date the data is. For example, if the system expects daily sales data but the latest batch is from three days ago, the freshness SLI would signal a problem.

Formally, freshness can be defined as: \[ \text{Freshness} = \text{Current Time} - \text{Timestamp of Latest Batch} \] Monitoring freshness ensures that analyses, reports, and dashboards are always using timely information. 

\item \textbf{Completeness:} Completeness evaluates whether all expected records or fields have been received. For instance, if a daily transaction report is missing a subset of stores or some fields are null, completeness would drop. Mathematically, this can be expressed as: \[ \text{Completeness} = \frac{\text{Number of Records Received}}{\text{Number of Records Expected}} \] A low completeness score indicates missing or partial data, prompting investigation before downstream processes rely on it. 

\item \textbf{Accuracy:} Accuracy measures how well the data complies with the validation rules defined in the data contract. For example, if ages, prices, or dates fall outside expected ranges, accuracy decreases. Formally: \[ \text{Accuracy} = 1 - \frac{\text{Number of Rule Violations}}{\text{Total Number of Records}} \] Maintaining high accuracy ensures that the data used for reporting and decision-making is trustworthy. 

\item \textbf{Contract Adherence:} This SLI checks whether every incoming batch respects the agreed-upon data contract, including both hard rules (which must be met) and soft rules (which may generate warnings). Contract adherence can be monitored as a percentage of batches fully compliant with the contract: \[ \text{Contract Adherence} = \frac{\text{Number of Compliant Batches}}{\text{Total Batches}} \] Tracking contract adherence provides visibility into whether upstream systems are delivering data in the expected format and structure. \end{itemize}

Once SLIs are defined, the pipeline is instrumented to automatically collect metadata and statistics for each batch, including timestamps, record counts, and validation errors. Quality scores are computed for each dataset and compared against pre-defined thresholds. If any metric falls below the acceptable level, automated alerts can notify engineers, or corrective actions can be triggered, such as re-ingestion, recomputation, or backfilling of missing data. Historical SLI logs are maintained to support reproducibility, auditing, and trend analysis, creating a feedback loop that continuously improves data quality over time.

\paragraph{Example: Daily Transaction Data Quality Template}

A concrete example illustrates how a template-based approach operationalizes these principles. Consider a daily transaction dataset. A reusable quality template could include the following checks:

\begin{itemize}
    \item \textbf{Freshness:} Verify that the latest batch timestamp is within 24 hours, using the freshness formula above.
    \item \textbf{Completeness:} Confirm that all expected stores and fields are present, as quantified by the completeness formula.
    \item \textbf{Accuracy:} Ensure that transaction amounts are non-negative and within expected ranges, monitored using the accuracy formula.
    \item \textbf{Contract Adherence:} Check that the schema matches the agreed definition and that required fields are present, measured with the contract adherence formula.
\end{itemize}

This template can be applied automatically to each daily batch. If any SLI falls below its threshold, alerts are triggered, and remediation actions—such as recomputation, data backfill, or manual review—can be executed. Over time, the consistent application of this template ensures that data entering downstream analytics is reliable, auditable, and actionable.

By embedding data quality directly into the pipeline as a template-based, reusable component, ETLT++ and ELTL++ transform quality assurance from a reactive, manual task into a proactive, automated, and scalable practice. This approach safeguards decision-making, maintains trust in analytics, and enables robust, reproducible processing of enterprise data.

Taken together, these enhancements make ETLT++ and ELTL++ reusable and systematic pattern for building pipelines that are not only flexible and auditable, but also governed, cost-effective, and user-friendly. This elevates ETLT++ and ELTL++ beyond an implementation practice into formal design patterns for modern data platforms, addressing recurring challenges in data engineering with solutions that are both practical and generalizable.

\section{Future Work}\label{futureworks}

This study opens several avenues for further research and refinement. We highlight four directions that appear most promising:

\begin{itemize}
    \item Empirical Evaluation: Implementing ETLT++ and ELTL++ in enterprise settings and benchmarking their performance against legacy ETL/ELT pipelines, with attention to latency, error rates, scalability, and audit readiness.
    \item Automated Tooling: Integrating the proposed design patterns with leading orchestration frameworks and cloud-native services in order to streamline deployment, testing, and monitoring.
    \item Expanded Usage Scenarios: Exploring adaptations of these patterns for streaming architectures, IoT workloads, and federated learning environments, where challenges of scale and heterogeneity are particularly acute.
    \item AI and ML Integration: Investigating how automated anomaly detection, adaptive contract enforcement, and predictive scaling can further enhance pipeline robustness through the application of machine learning techniques.
\end{itemize}

\subsection{Evaluation Criteria and Benchmark Plan}

An essential direction for future work is the empirical validation of ETLT++ and ELTL++. While the present contribution has focused on conceptual formalization and enhanced design principles, systematic benchmarking is necessary to demonstrate effectiveness in practice. Conducting such evaluations across multiple cloud environments (e.g., Snowflake, BigQuery, Databricks) would require significant engineering resources and is therefore beyond the scope of this paper. Nevertheless, we propose a feasible benchmark plan that can be initially carried out in a controlled environment and subsequently extended to larger contexts.

The purpose of this benchmark is not to exhaustively compare every technological stack, but rather to illustrate the value of ETLT++ and ELTL++ in addressing recurring challenges in modern data pipelines. We identify five evaluation dimensions that capture their distinctive properties:
\begin{itemize}
    \item Latency and throughput: time required for data to move from ingestion to analytical consumption.
    \item Error containment: proportion of invalid or contract-violating records intercepted at T1 in ETLT++, or flagged through monitoring templates in ELTL++.
    \item Reproducibility and auditability: ability to replay historical queries with consistent results using versioned raw storage.
    \item Cost efficiency: balance between storage overhead for raw and curated layers and compute resources used for transformations.
    \item Operational resilience: mean recovery time following upstream schema changes or data quality violations.
\end{itemize}

A lightweight validation of these criteria can be conducted using a single open dataset, such as the NYC Taxi Trips or a representative IoT event stream. The dataset may be processed under two configurations within the same environment: (i) a baseline ELT pipeline, and (ii) an enhanced ETLT++ or ELTL++ pipeline implemented with open-source lakehouse technologies such as Apache Spark and Delta Lake or Apache Iceberg. Restricting the setup to a single environment avoids the complexity of multi-cloud benchmarking while still yielding meaningful comparisons across the five dimensions.

Pipeline observability and data quality can be evaluated using open-source frameworks such as Great Expectations \cite{greatexpectations} or Deequ \cite{schelter2018deequa}, which provide standardized checks for schema conformance, missing values, and value ranges. These tools make it possible to systematically measure error containment and contract adherence without bespoke validation infrastructure.

The ambition of this benchmark plan is not to provide immediate large-scale empirical proof, but to establish a structured roadmap for validation. Future research should extend the approach across multiple cloud platforms and industrial-scale datasets, thereby quantifying trade-offs in latency, cost, and resilience under heterogeneous operational contexts. In this way, the proposed evaluation plan provides both a pragmatic first step and a foundation for the systematic assessment of enhanced hybrid patterns in modern data engineering.

\section{Conclusion}

This work formalized ETLT and ELTL as advanced, hybrid design patterns for modern data engineering and introduced their enhanced forms, ETLT++ and ELTL++. By systematically separating data quality controls from business logic, versioning raw data, and embedding continuous monitoring, these patterns provide a reusable blueprint for efficient, auditable, and resilient data pipelines across diverse, multi-cloud environments.

The proposed patterns address critical challenges, such as quality assurance, lineage, and multi-team usage, that traditional ETL and ELT approaches alone cannot resolve. Through enforced data contracts, append-only loading, standardized transformation templates, and automated SLI/SLO monitoring, ETLT++ and ELTL++ promise measurable improvements in operational reliability, governance, and agility. These contributions equip practitioners with practical tools to build scalable platforms that meet both regulatory and analytical requirements.

Unlike prior work, this paper is the first to systematically define ETLT and ELTL as formal design patterns, extend them with enforceable obligations (contracts, versioning, monitoring), and provide reusable templates for operational adoption. Crucially, ETLT++ and ELTL++ should not be interpreted as incremental refinements of ETL/ELT, but as formally defined design patterns that embed contracts, versioning, and observability as obligations, thus transforming best practices into systematic architectural guarantees.

\section{Acknowledgements}

\section*{Declarations}

\begin{itemize}
\item Data availability 

This study did not use any publicly available or proprietary datasets. All data referenced in the article consist of dummy data provided in the text for illustrative purposes only and do not contain any sensitive information. No additional datasets or materials were generated or analyzed.
\end{itemize}

\printbibliography
\end{document}